\documentclass[a4paper]{aa}

\usepackage{graphicx,natbib,amsmath}
\usepackage{txfonts}
\usepackage[usenames,dvipsnames]{xcolor}

\newcommand{\sco}{$\tau$~Sco}

\newcommand{\vs}{$v_{\rm e}\sin i$}
\newcommand{\kms}{km\,s$^{-1}$}

\newcommand{\figps}[1]{\resizebox{\hsize}{!}{\rotatebox{0}{\includegraphics{#1}}}}

\newcommand{\fifps}[2]{\centering\resizebox{#1}{!}{\includegraphics{#2}}}

\newcommand{\beq}{\begin{equation}}
\newcommand{\eeq}{\end{equation}}

\begin{document}

\title{Magnetic field topology of $\tau$~Sco%
\thanks{Based on observations obtained at the Canada-France-Hawaii Telescope (CFHT) which is operated by the National Research Council of Canada, the Institut National des Sciences de l'Univers of the Centre National de la Recherche Scientifique of France, and the University of Hawaii.}}
\subtitle{The uniqueness problem of Stokes $V$ ZDI inversions}

\author{O.~Kochukhov\inst{1}
  \and G.~A.~Wade\inst{2}
}

\institute{
Department of Physics and Astronomy, Uppsala University, Box 516, 75120 Uppsala, Sweden
\and
Department of Physics, Royal Military College of Canada, PO Box 17000, Station `Forces', Kingston, ON K7K 7B4, Canada
}

\date{Received 25 September 2015 / Accepted 24 November 2015}

\titlerunning{Magnetic field topology of $\tau$~Sco}
\authorrunning{O. Kochukhov and G. A. Wade}

\abstract
{
The early B-type star \sco\ exhibits an unusually complex, relatively weak surface magnetic field. Its topology was previously studied with the Zeeman Doppler imaging (ZDI) modelling of high-resolution circular polarisation (Stokes $V$) observations.
}
{
Here we assess the robustness of the Stokes $V$ ZDI reconstruction of the magnetic field geometry of \sco\ and explore the consequences of using different parameterisations of the surface magnetic maps. 
}
{
This analysis is based on the archival ESPaDOnS high-resolution Stokes $V$ observations and employs an independent ZDI magnetic inversion code.
}
{
We succeeded in reproducing previously published magnetic field maps of \sco\ using both general harmonic expansion and a direct, pixel-based representation of the magnetic field. These maps suggest that the field topology of \sco\ is comprised of comparable contributions of the poloidal and toroidal magnetic components. At the same time, we also found that available Stokes $V$ observations can be successfully fitted employing restricted harmonic expansions, by either neglecting the toroidal field altogether or linking the radial and horizontal components of the poloidal field as required by the widely used potential field extrapolation technique. These alternative modelling approaches lead to a stronger and topologically more complex surface field structure. The  field distributions recovered with different ZDI options differ significantly, yielding indistinguishable Stokes $V$ profiles but different linear polarisation (Stokes $Q$ and $U$) signatures.
}
{
Our investigation underscores the well-known problem of non-uniqueness of the Stokes $V$ ZDI inversions. For the magnetic stars with properties similar to \sco\ (relatively complex field, slow rotation) the outcome of magnetic reconstruction depends sensitively on the adopted field parameterisation, rendering photospheric magnetic mapping and determination of the extended magnetospheric field topology ambiguous. Stokes $Q$ and $U$ spectropolarimetric observations represent the only way of breaking the degeneracy of surface magnetic field models. 
}


\maketitle

\section{Introduction}
\label{intro}

Tomographic inversion of spectral timeseries observations of stars (referred to as ``Doppler imaging'') is an important modern tool for the characterisation of stellar surface structure. Since the introduction of these methods in the 1970s to reconstruct the chemical abundance spot distributions of Ap stars \citep{khokhlova:1976}, they have been extended to map temperature inhomogeneities of cool stars \citep[e.g.][]{vogt:1983b}, surface velocity fields of pulsating stars \citep{kochukhov:2004f}, and stellar surface magnetic fields \citep{donati:1990,brown:1991,piskunov:2002a} (the latter with the acquisition of spectropolarimetric, rather than spectral, timeseries).

The basic principles underlying Doppler imaging approaches are rotational modulation in combination with indirect (i.e. spectral) resolution of the stellar surface. The latter is achieved through exploitation of the rotational Doppler broadening of spectral line profiles, in combination with the temporal evolution of the profile. 

Because Doppler imaging relies on the inversion of a spectral timeseries to reconstruct a 2D surface map, it is inherently ill-posed in most of its implementations. As a consequence, significant efforts have been invested to understand the accuracy and uniqueness of the derived maps \citep[e.g.][]{rice:1991a,rice:2000}. In the reconstruction of stellar magnetic fields (known as ``Zeeman Doppler imaging'', or ZDI), an important theme has been to understand the uniqueness of maps obtained from spectropolarimetric timeseries in which the Stokes vector is not fully characterised \citep[e.g.][]{donati:1997a,kochukhov:2002c,rosen:2012}. This problem potentially affects most magnetic maps reconstructed using ZDI, as the large majority are inferred from inversion of only Stokes $I$ and $V$ spectra, i.e. without knowledge of the Stokes $Q$ and $U$ linear polarisation parameters.

\citet{donati:2006b}, hereafter D06, reported the discovery of a medium-strength ($\sim$\,0.5~kG) magnetic field on the young, massive star $\tau$~Sco (B0.2 V, HR~6165, HD~149438). Circularly polarised Zeeman signatures, modulated according to the $\sim$\,41~d rotational period of the star, were clearly detected in observations collected mostly with the ESPaDOnS spectropolarimeter. Using a ZDI approach, D06 reconstructed the large-scale structure of the magnetic topology of \sco, finding a magnetic structure that was unusually complex for a hot star, with significant power in spherical-harmonic modes up to degree 5. The surface topology they recovered is dominated by a potential field, although they concluded that a moderate toroidal component is probably also present. They also extrapolated the reconstructed surface magnetic field into the region above the stellar surface using a potential, force-free assumption. The topology of the extended magnetic field they derived is also more complex than a global dipole, and features in particular a significantly warped torus of closed magnetic loops encircling the star, with additional, smaller, networks of closed field lines. \citet{donati:2009} published an updated model of the surface field using a more extensive data set.

\citet{ignace:2010} observed $\tau$ Sco using the Suzaku X-ray observatory. Based on the potential field extrapolation of D06, two main X-ray eclipses were expected at phases of around 0.3 and 0.8, in conjunction with enhanced UV line absorptions observed at those times (D06). While the Suzaku pointings  include phases close to those at which the eclipses were predicted, no X-ray variations anywhere near the expected amplitudes were detected. 

The lack of X-ray eclipses likely implies that the X-ray formation region is located significantly further away from the star than is predicted according to the ZDI reconstruction and potential extrapolation of D06. 

To better understand the origin of this discrepancy, we have investigated uniqueness of the surface magnetic maps of $\tau$~Sco derived from the Stokes $V$ timeseries of D06. In particular, we have assessed sensitivity of the ZDI reconstruction to the choice of parameterisation of the stellar surface magnetic field distribution. 

The rest of the paper is organised as follows. In Sect.~\ref{methods} we discuss spectropolarimetric observations of \sco\ and summarise alternative approaches to parameterisation of the surface magnetic field in ZDI. Sect.~\ref{results} presents magnetic maps obtained using different surface field parameterisations and assesses the consequence of these differences for the magnetic energy distribution at different spatial scales and for the magnetospheric field structure. The results of our study are summarised and discussed in Sect.~\ref{concl}.

\section{Methods}
\label{methods}

\subsection{Spectropolarimetric observational data}

We employed the same collection of CFHT ESPaDOnS observations of $\tau$~Sco as was used by \citet{donati:2009}, apart from the addition of a single additional observation acquired in July 2009. This observation corresponds to phase 0.740 (a phase at which an earlier observation already exists).

All observations were extracted from the CFHT Archive in reduced form, and normalised to the continuum order-by-order using polynomial fits. Least-Squares Deconvolution (LSD) was applied to each spectrum using the {\sc iLSD} procedure as implemented by \citet{kochukhov:2010a}. We used a line mask based on a Vienna Atomic Line Database \citep[{\sc vald},][]{kupka:1999} ``Extract Stellar'' request, retaining only lines stronger than 10\% of the continuum, cleaned and adjusted in depth to best match the observed spectrum of $\tau$~Sco.

All extracted profiles are scaled corresponding to a Land\'e factor of 1.20, an unbroadened central depth of 0.2, and a wavelength of 500~nm. Observations from the same nights were averaged and one low signal-to-noise ratio spectrum was excluded, resulting in a data set comprising 49 rotational phases.

\subsection{Parameterisation of magnetic maps in ZDI}

\subsubsection{Direct parameterisation}

Early Zeeman Doppler imaging studies \citep{brown:1991,piskunov:2002a} treated each component of the surface magnetic field vector distribution as an independent two-dimensional map defined on a discrete longitude-latitude grid. Each of the three magnetic field component maps were simultaneously fitted to spectropolarimetric timeseries data, employing one or another form of regularisation (e.g. maximum entropy or Tikhonov methods). This approach to the stellar magnetic field mapping, hereafter called \textit{direct inversion}, results in magnetic field distributions which do not necessarily satisfy  Maxwell's equations. In particular, the net signed flux through the stellar surface can significantly deviate from zero. 

Other, less obvious problems of the direct field parameterisation approach include various degeneracies and cross-talks between different magnetic field components when ZDI is based only on the Stokes $V$ spectra \citep{brown:1991,kochukhov:2002c,rosen:2012}. Moreover, the choice of regularisation applied in the direct ZDI problem appears to be critical for adequate reconstruction of certain types of globally-organised magnetic field topologies. For instance, according to \citet{brown:1991} and \citet{donati:2001a} direct ZDI guided by the maximum entropy constraint fails to recover a simple dipolar field configuration, even using spectra in all four Stokes parameters. Conversely, the ZDI employing Tikhonov regularisation succeeds in reconstructing dipolar and low-order multipolar fields \citep{piskunov:2002a,kochukhov:2002c}.

\subsubsection{General spherical-harmonic parameterisation}

Spherical-harmonic expansion is an alternative form of the stellar magnetic field parameterisation favoured by the more recent ZDI studies \citep{donati:2001a,donati:2006b,kochukhov:2014}. In this method the free parameters of the magnetic inversion problem are coefficients specifying amplitudes of various spherical-harmonic modes. The most general form of the harmonic representation of an arbitrary surface vector field distribution is given by
\beq
B_{\rm r} (\theta, \phi) = -\sum_{\ell=1}^{\ell_{\rm max}} \sum_{m=-\ell}^{\ell} \alpha_{\ell,m} Y_{\ell,m} (\theta, \phi),
\label{eq1}
\eeq
\beq
B_{\theta} (\theta, \phi) = -\sum_{\ell=1}^{\ell_{\rm max}} \sum_{m=-\ell}^{\ell} \left[ \beta_{\ell,m} Z_{\ell,m} (\theta, \phi) + \gamma_{\ell,m} X_{\ell,m} (\theta,\phi) \right],
\label{eq2}
\eeq
\beq
B_{\phi} (\theta, \phi) = -\sum_{\ell=1}^{\ell_{\rm max}} \sum_{m=-\ell}^{\ell}\left[ \beta_{\ell,m} X_{\ell,m} (\theta, \phi) - \gamma_{\ell,m} Z_{\ell,m} (\theta,\phi) \right],
\label{eq3}
\eeq
where $Y_{\ell,m} (\theta, \phi)$, $Z_{\ell,m} (\theta, \phi)$ and $X_{\ell,m} (\theta, \phi)$ are spherical-harmonic functions of the angular degree $\ell$ and azimuthal order $m$ and their derivatives with respect to the latitude $\theta$ and longitude $\phi$ (see \citet{kochukhov:2014} for definitions of these quantities). The coefficients $\alpha_{\ell,m}$ and $\beta_{\ell,m}$ characterise, respectively, the vertical and horizontal components of the poloidal (potential) field. The $\gamma_{\ell,m}$ coefficients specify the strength of the toroidal (non-potential) field components.

It is easy to see several attractive features of the ZDI based on the harmonic field description. This representation of the field structure automatically obeys the requirement of zero total magnetic flux piercing through the closed surface (one of Maxwell's equations). It also enables detailed quantitative characterisation of ZDI results by making it straightforward to, e.g., assess the strength of the poloidal vs. toroidal field components or axisymmetric vs. non-axisymmetric field. It was also claimed by \citet{donati:2001a} that the spherical-harmonic parameterisation alleviates to some extent the problem of the radial/meridional field cross-talk in the ZDI with Stokes $V$ data alone.

\subsubsection{Constrained spherical-harmonic parameterisation}
\label{pssf:intro}

The description of the surface magnetic field geometry in terms of the \textit{general harmonic expansion} given by Eqs.~(\ref{eq1})--(\ref{eq3}) carries along a large number of degrees of freedom, which may be superfluous for particular ZDI problems. For example, the necessity of including the toroidal field components given by the coefficients $\gamma_{\ell,m}$ is not always explicitly justified by ZDI studies. On the other hand, using independent sets of the $\alpha_{\ell,m}$ and $\beta_{\ell,m}$ coefficients implies independent treatment of the horizontal and vertical poloidal field components, which appears to be unnecessary e.g. in the context of studies of the magnetic field topologies of early-type stars \citep{landstreet:2000,bagnulo:2002}.

In addition to simplicity arguments, there are fundamental physical reasons to expect a link between $\alpha_{\ell,m}$ and $\beta_{\ell,m}$. The method of potential source surface field (PSSF) extrapolation \citep{jardine:2002a,jardine:2013} is frequently used to analyse the extended magnetospheric field topology based on the photospheric ZDI map of the radial field component. It was applied to \sco\ by D06. This technique imposes a relation between $\alpha_{\ell,m}$ and $\beta_{\ell,m}$ for a given location of the source surface $R_{\rm s}$. For example, with the definitions of \citet{kochukhov:2013} and the stellar radius $R_\star$, one can determine
\beq
\beta_{\ell,m}=\alpha_{\ell,m} \dfrac{(\ell+1) (R_{\rm s}/R_{\star})^{2\ell+1} - \ell -1}{(\ell +1) (R_{\rm s}/R_{\star})^{2\ell+1} + \ell},
\label{eq4}
\eeq
which reduces to the condition $\beta_{\ell,m}\approx\alpha_{\ell,m}$ when $R_{\rm s} \gg R_\star$.

Somewhat surprisingly, despite a wide-spread application of the PSSF extrapolations of ZDI results, almost all magnetic mapping studies \citep[with the exception of][]{hussain:2002} do not attempt to incorporate this constraint in the magnetic inversions themselves. As discussed by \citet{jardine:2013}, this leads to an inconsistency between the poloidal horizontal field reconstructed by ZDI and the horizontal field calculated with the PSSF technique. The ZDI maps of \sco\ and corresponding field extrapolation presented by D06 also suffer from this inconsistency.

\subsection{Application of magnetic inversions to \sco}

The main goal of our study is to explore sensitivity of the ZDI modelling of \sco\ to different forms of magnetic field parameterisation. To this end, we have carried out Stokes $V$ magnetic inversions using the direct field parameterisation (Model 1), the general harmonic parameterisation ($\gamma_{\ell,m}\ne0$, $\beta_{\ell,m}\ne\alpha_{\ell,m}$, Model 2), and two constrained harmonic field representations. Specifically, we tested feasibility of fitting the circular polarisation observations of \sco\ with a purely poloidal field ($\gamma_{\ell,m}=0$, $\beta_{\ell,m}\ne\alpha_{\ell,m}$, Model 3) and with a mixture of toroidal and constrained poloidal fields ($\gamma_{\ell,m}\ne0$, $\beta_{\ell,m}=\alpha_{\ell,m}$, Model 4). For the relatively complex field of \sco\ and its source surface radius of $R_{\rm s}$\,=\,1.8--2.0$R_\star$ \citep[D06,][]{petit:2013} the latter parameterisation is essentially equivalent to application of Eq.~(\ref{eq4}). In all inversions discussed below the spherical-harmonic expansion was truncated to $\ell=10$. Similar to D06, we found that none of the models with $\ell \le 2$ can produce an acceptable fit to the data.

Reconstruction of the magnetic field topology of \sco\ was carried out the help of the {\sc InversLSD} ZDI code described by \citet{kochukhov:2013}. The Tikhonov regularisation \citep{tikhonov:1977} was applied for the direct inversion (Model 1); a penalty function minimising the energy of high-order modes \citep{kochukhov:2013} was employed to constrain the harmonic inversions (Models 2--4). Since no detectable variability is seen in Stokes $I$, only the Stokes $V$ LSD profiles were modelled. 

We adopted a projected rotational velocity of \vs\,=\,6~\kms\ and an inclination angle of $i=70\degr$ following D06. The local Stokes $V$ profile shape was approximated with the Unno-Rachkovsky analytical solution of the polarised radiative transfer equation in the Milne-Eddington atmosphere \citep{polarization:2004}. For these calculations the Zeeman splitting pattern of the average line was represented by a triplet with the same effective Land\'e factor $g_0=1.2$ and wavelength $\lambda_0=500$~nm as were adopted for normalisation of the observed LSD profiles. The depth and width of the theoretical local profile were adjusted to match the observed Stokes $I$ LSD line shape.

\section{Results}
\label{results}

\begin{table*}[!th]
\centering
\caption{Summary of ZDI inversions for \sco. 
\label{tbl:zdi}}
\begin{tabular}{ccc|ccc|ccc|c}
\hline\hline
Model   & Parameterisation & $\sigma_V$ & \multicolumn{3}{c|}{RMS field (G)}  & \multicolumn{3}{c|}{Magnetic energy} & Plot \\
number & of magnetic maps  & $\times 10^{-5}$ & $\langle B_{\rm r} \rangle$ & $\langle B_{\theta} \rangle$ & $\langle B_{\varphi} \rangle$ & total & pol:tor & $\ell < m/2$:$\ell \ge m/2$ & \\
\hline
1 & direct & 7.9 & 168 & 142 & 211 & 1.18 & -- & -- & Fig.~\ref{fig:map1}a \\
2 & harmonic, $\beta_{\ell,m} \ne \alpha_{\ell,m}, \gamma_{\ell,m} \ne 0$ & 8.1 & 160 & 118 & 215 & 1.09 & 45.6:54.4 & 25.6:74.4 & Fig.~\ref{fig:map1}b \\
3 & harmonic, $\beta_{\ell,m} \ne \alpha_{\ell,m}, \gamma_{\ell,m} = 0$ & 8.2 & 356 & 206 & 223 & 2.76 & 100.0:0.0 & 34.3:65.7 & Fig.~\ref{fig:map2}a \\
4 & harmonic, $\beta_{\ell,m} = \alpha_{\ell,m}, \gamma_{\ell,m} \ne 0$ & 8.4 & 663 & 900 & 598 & 20.34 & 48.6:51.4 & 6.3:93.7 & Fig.~\ref{fig:map2}b \\
\hline
\end{tabular}
\tablefoot{The columns give ZDI inversion number, assumed parameterisation of magnetic field maps, mean deviation of the fit to observed Stokes $V$ profiles, rms values of the radial, meridional, and azimuthal field components, the total magnetic energy (in arbitrary units), and the relative energies contained in the poloidal vs. toroidal and axisymmetric vs. non-axisymmetric field components.}
\end{table*}

\begin{figure*}[!th]
\centering
\fifps{8.9cm}{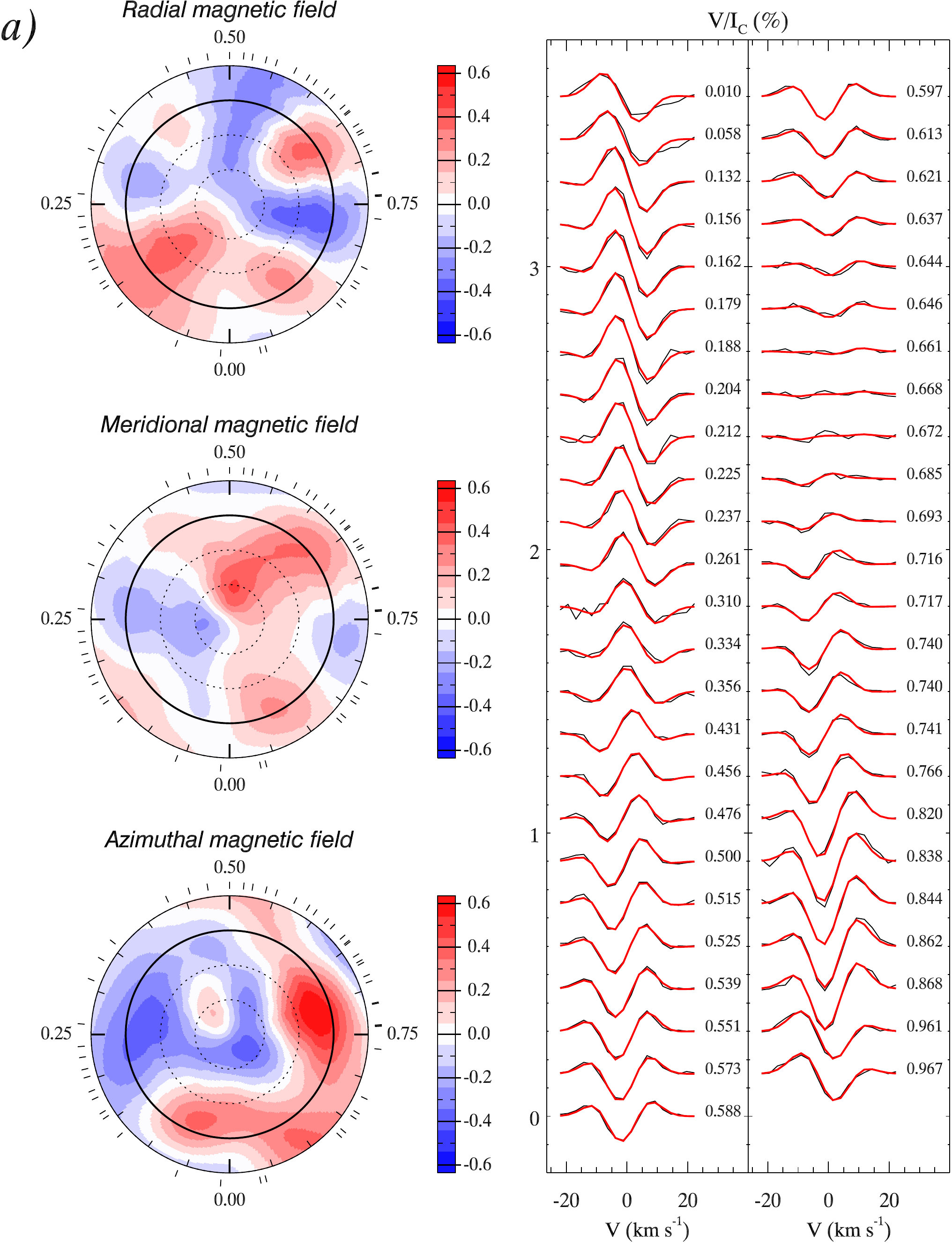}\hspace*{0.5cm}
\fifps{8.9cm}{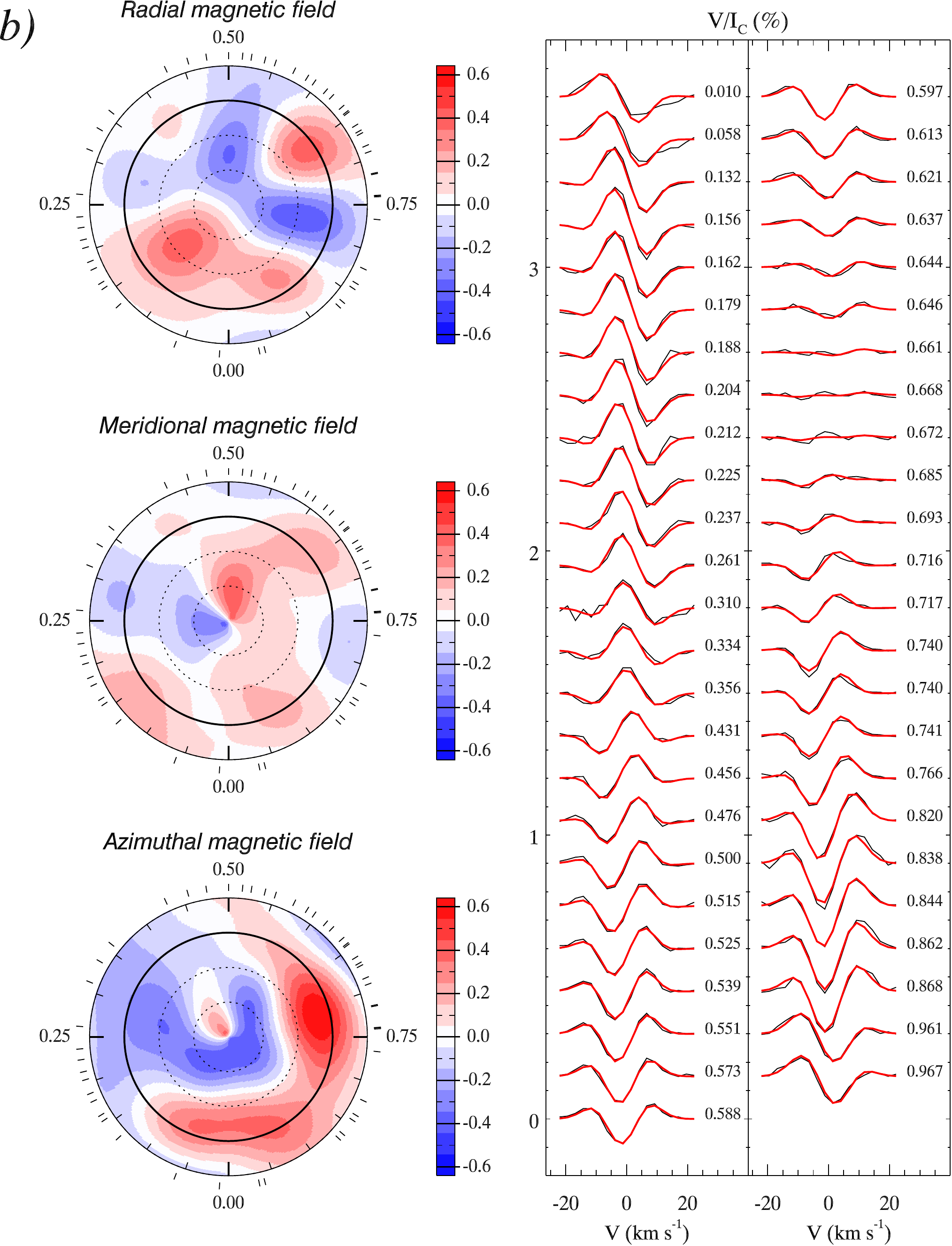}
\caption{Global magnetic field topology of \sco\ derived with ZDI inversions for the case of {\bf a)} direct parameterisation of the magnetic maps (Model 1) and {\bf b)} harmonic parameterisation with $\beta_{\ell,m} \ne \alpha_{\ell,m}, \gamma_{\ell,m} \ne 0$ (Model 2). In each panel the left column shows the flattened polar projection of the radial, meridional and azimuthal field components. The colour bar indicates the field strength in kG. The observed (thin black line) and computed (thick red line) LSD Stokes $V$ profiles are compared on the right side of each panel. The profiles are offset vertically according to the rotational phase indicated next to each spectrum.}
\label{fig:map1}
\end{figure*}

\begin{figure*}[!th]
\centering
\fifps{8.9cm}{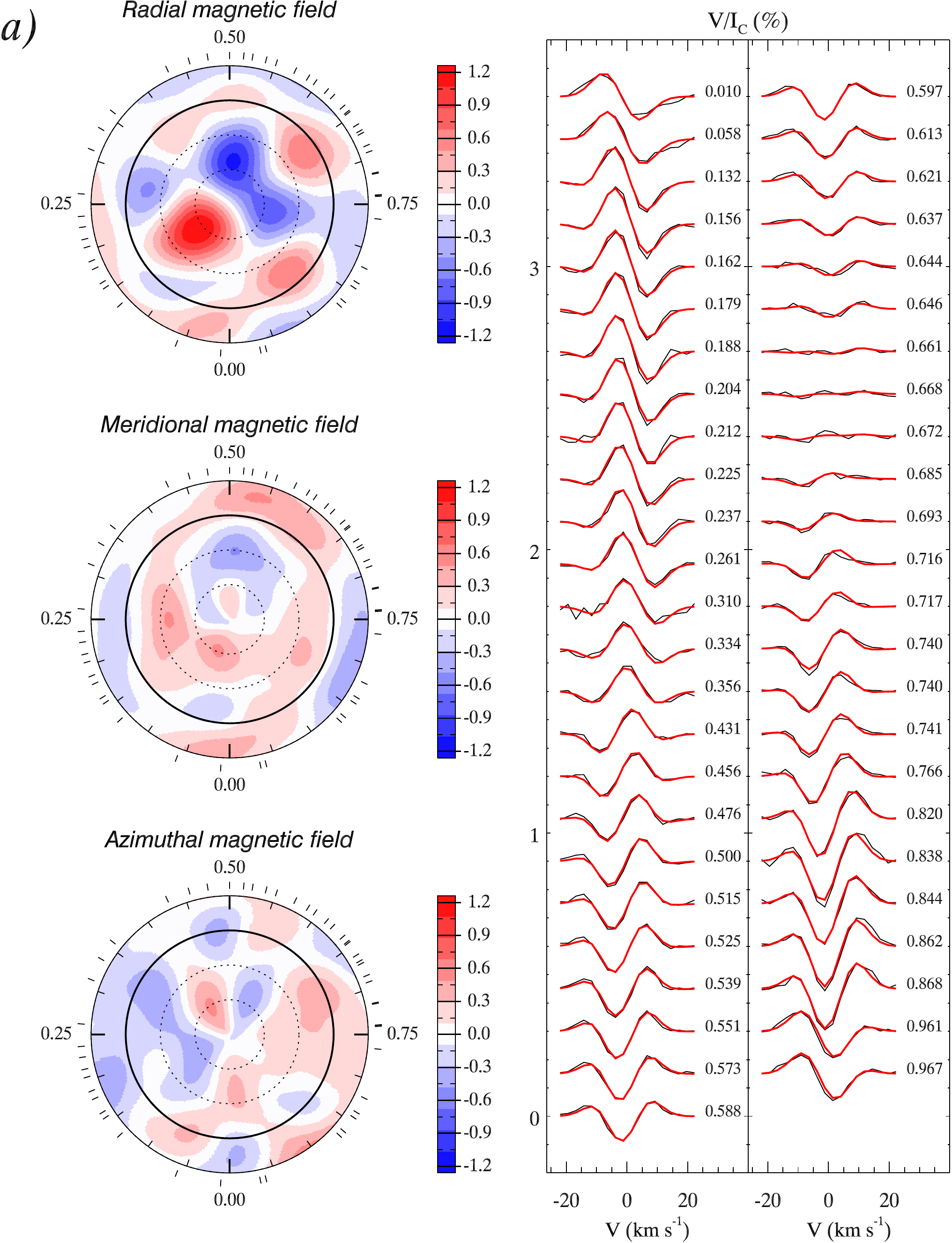}\hspace*{0.5cm}
\fifps{8.9cm}{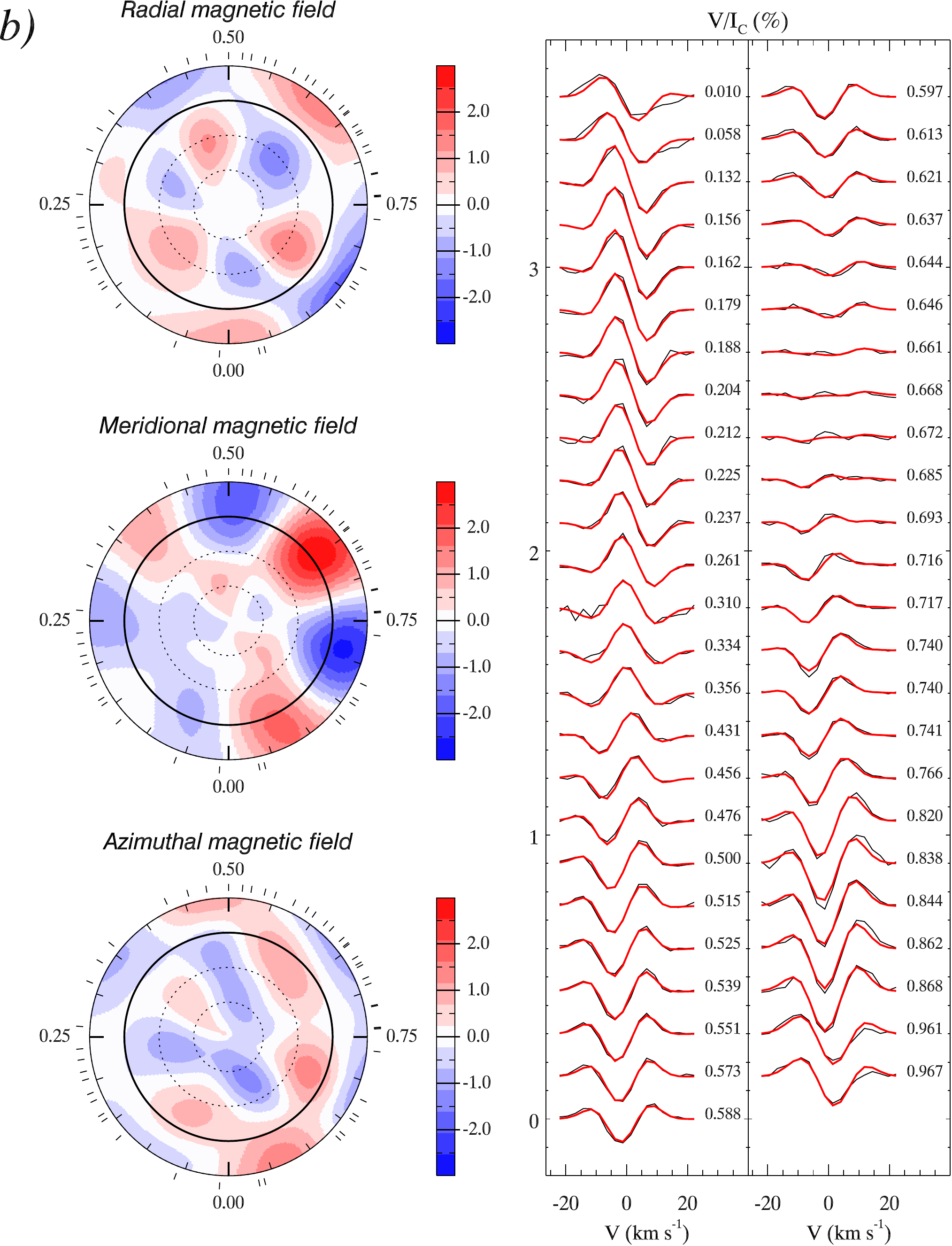}
\caption{Same as Fig.~\ref{fig:map1} for the harmonic ZDI inversions with {\bf a)} $\beta_{\ell,m} \ne \alpha_{\ell,m}, \gamma_{\ell,m} = 0$ (Model 3) and {\bf b)} $\beta_{\ell,m} = \alpha_{\ell,m}, \gamma_{\ell,m} \ne 0$ (Model 4).}
\label{fig:map2}
\end{figure*}

\subsection{Magnetic field distributions}

Results of the ZDI inversions based on different surface field parameterisations are summarised in Table~\ref{tbl:zdi}. For each inversion, this table gives a measure of the quality of the final fit to the observed Stokes $V$ LSD profiles, the rms values of the three magnetic field components, the total magnetic energy and an estimate of the relative energy contributions of the poloidal/toroidal and axisymmetric/non-axisymmetric field components.

The resulting magnetic field maps and corresponding line profile fits are presented in Figs.~\ref{fig:map1} and \ref{fig:map2}. For each ZDI model these figures illustrate the radial, meridional and azimuthal field distributions over the surface of \sco\ in the flattened polar projection, similar to the presentation used by D06.

The inversions corresponding to Model~1 (direct ZDI) and 2 (general spherical-harmonic ZDI) are shown in Fig.~\ref{fig:map1}. Evidently, these magnetic field maps are very similar in all three magnetic vector components. There is no evidence that reliance on a spherical-harmonic expansion leads to an appreciably different field distribution, despite an additional physical constraint. We conclude that for \sco\ the effective degrees of freedom in the direct and harmonic inversions are largely equivalent.

A comparison of Fig.~\ref{fig:map1} with the magnetic field distribution published by D06 and with an updated magnetic field map of \sco\ presented by \citet{donati:2009} shows an excellent agreement. Both details of the surface distributions of the individual field components and their amplitudes agree very well. Very small remaining differences may be related to a different local Stokes $V$ profile shape adopted in the two studies.

\begin{figure}[!th]
\centering
\fifps{4.3cm}{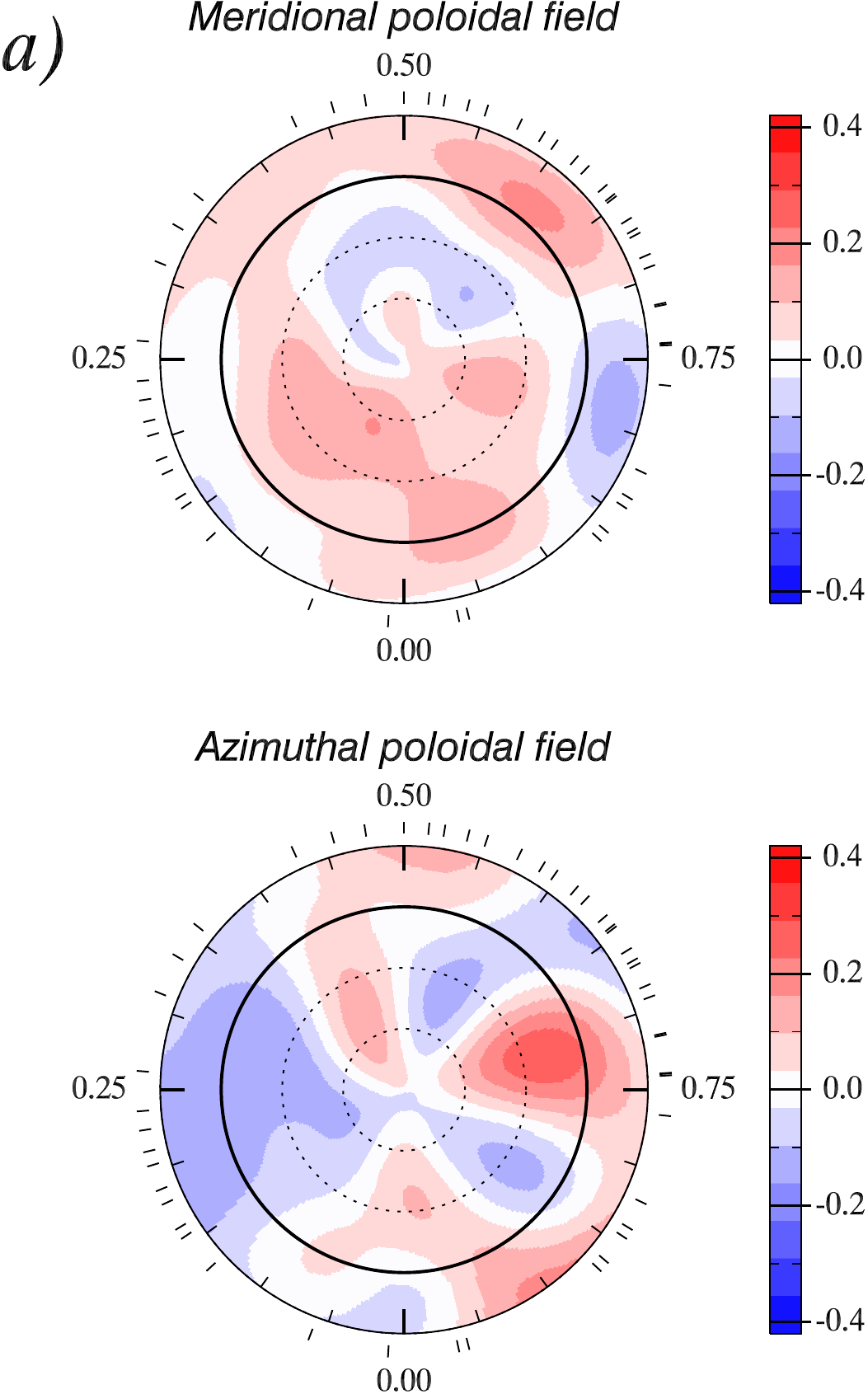}\hspace*{0.1cm}
\fifps{4.3cm}{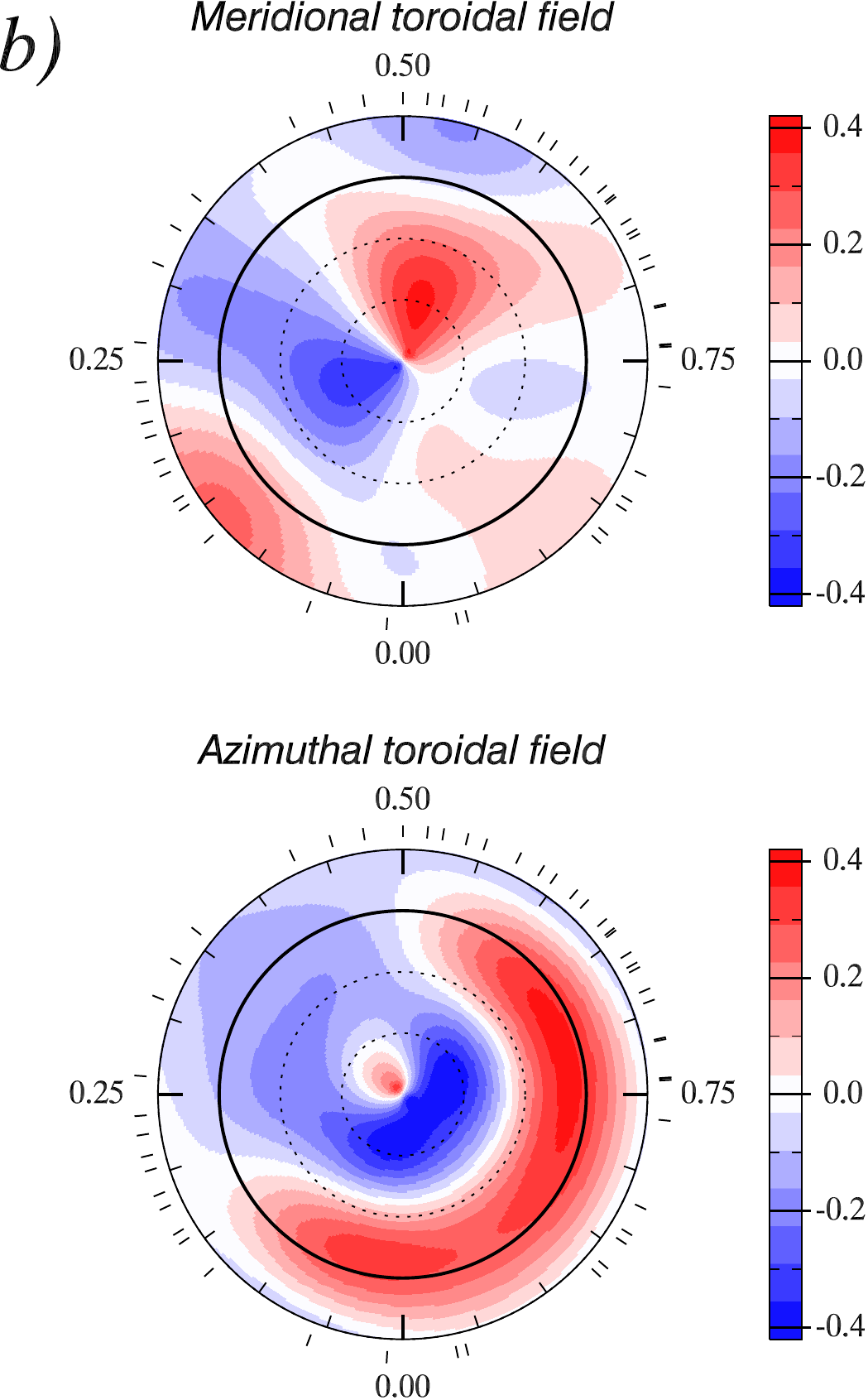}
\caption{Comparison of the {\bf a)} poloidal and {\bf b)} toroidal contributions to the horizontal magnetic field obtained in the general harmonic ZDI inversion (Model 2) illustrated in Fig.~\ref{fig:map1}b.}
\label{fig:map1a}
\end{figure}

Despite this encouraging concordance of the magnetic maps themselves, their interpretation is somewhat different. D06 reported that the potential field contribution clearly dominates, in terms of the magnetic energy, over the toroidal field, although both are required to fit the observations. Instead, we found that the toroidal field is slightly stronger than the poloidal contribution. The importance of the toroidal field is confirmed by Fig.~\ref{fig:map1a}, which presents the horizontal field maps separately for the poloidal and toroidal components of the field distribution corresponding to Model~2. It is evident that the non-potential field components dominate for both the meridional and azimuthal field maps.

We cannot offer a definitive explanation for the discrepancy in the interpretation of essentially the same magnetic map obtained here and by D06. 
In our analysis the energy of each spherical-harmonic field component was established by direct integration of the corresponding $B_{\ell,m}^2(\theta,\phi)$ maps over the stellar surface. The same approach was used by D06 (P. Petit, private communication). We have verified that restricting the magnetic inversion to the same rotational phases as used by D06 does not change the inferred relative magnetic energy contributions of the poloidal and toroidal field components.

Results of the ZDI with constrained spherical-harmonic expansions are presented in Fig.~\ref{fig:map2}. Both Model~3 and 4 yield an acceptable fit to the observed Stokes $V$ profiles, which is only marginally worse than the fits achieved by Models 1 and 2. In comparison, D06 reported a significantly worse fit with a purely poloidal field topology than with a poloidal plus toroidal field configuration. These authors did not consider the constrained harmonic field parameterisation equivalent to our Model 4.

In our inversions a good fit with constrained harmonic expansions is achieved at the expense of making the field stronger and more complex. In particular, the ZDI inversion ignoring toroidal field (Model~3) recovers local field strengths of up to 1.2~kG and has the total energy 2.3--2.5 times that of Models 1 and 2. The radial field distribution seen in Fig.~\ref{fig:map2}a is somewhat reminiscent of the field structure shown in Figs.~\ref{fig:map1}. At the same time, the horizontal field topology is entirely different.

Still stronger field is required to obtain an adequate description of observations when the toroidal field is present but the horizontal and radial poloidal fields are assumed to be linked ($\alpha_{\ell,m}=\beta_{\ell,m}$, Model~4). Both the radial and horizontal components of the resulting field distribution are significantly different from those obtained before; the maximum local surface field strength exceeds 2.5~kG and the total magnetic energy is now almost 20 times larger compared to Models~1 and 2.

\begin{figure}[!th]
\centering
\fifps{7cm}{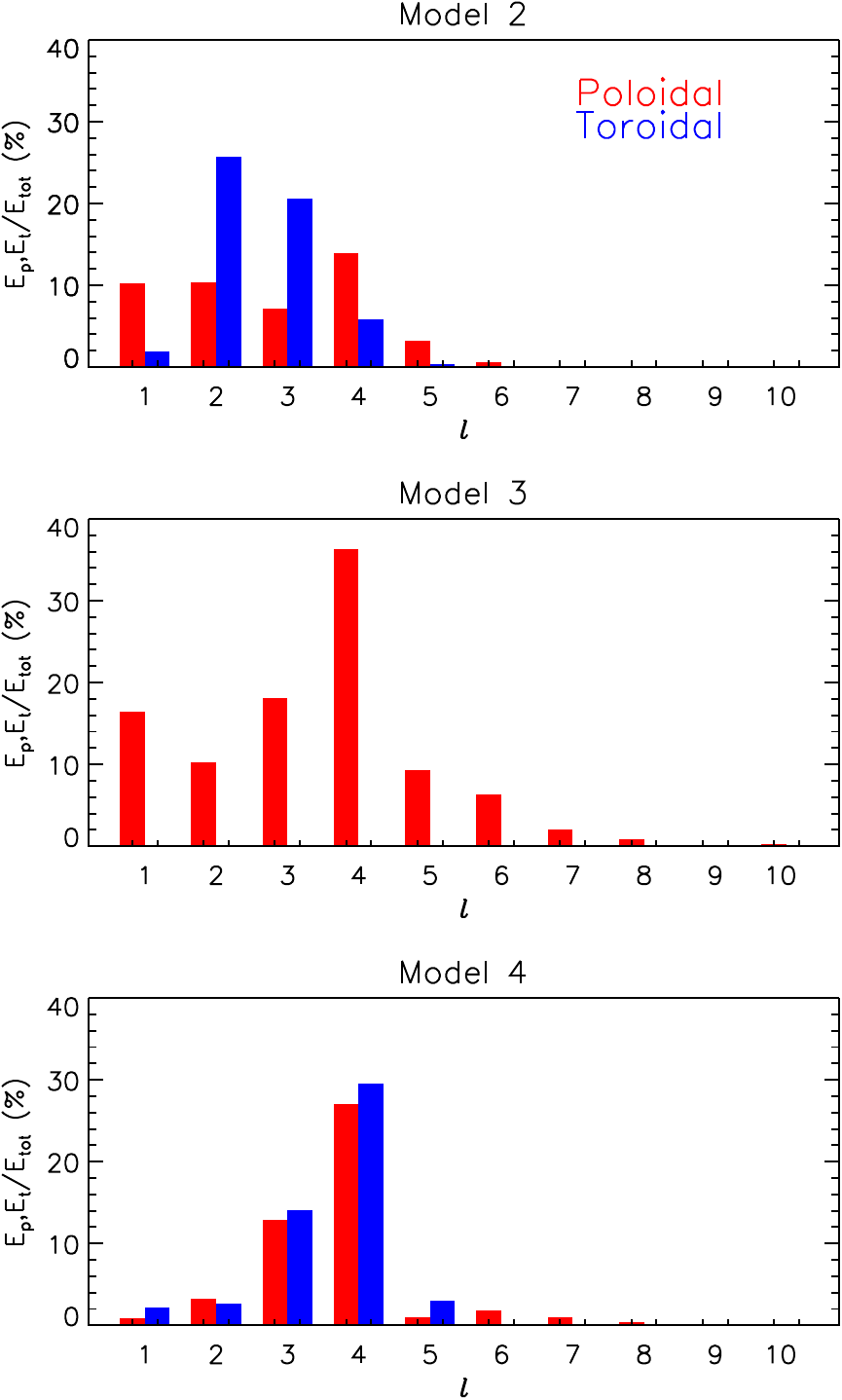}
\caption{Distribution of magnetic energy over different harmonic modes for ZDI inversions 2--4. The energy of poloidal components (light bars) is compared with toroidal contribution (dark bars) at different angular degree $\ell$.}
\label{fig:energ}
\end{figure}

\subsection{Magnetic energy spectrum}

In addition to the assessment of the total poloidal/toroidal component contributions, reported in Table~\ref{tbl:zdi}, we have analysed a distribution of the magnetic field energy over different $\ell$-modes for all inversions based on the spherical-harmonic expansion. The fractional energy in different poloidal and toroidal modes is illustrated in Fig.~\ref{fig:energ}. For Model 2 (general harmonic parameterisation), the poloidal field is spread over $\ell$\,=\,1--4 while the toroidal field peaks at $\ell$\,=\,2--3. For Model 3 (purely poloidal field), the energies are spread over the range $\ell$\,=\,1--6, with $\ell=4$ providing the maximum contribution. For Model~4 (linked horizontal and vertical components of the poloidal field), both the poloidal and toroidal fields are dominated by the contribution from $\ell$\,=\,3--4 modes. The lowest order, dipolar, contribution is also much smaller for Model~4 compared to other ZDI inversions. To summarise, the spatial magnetic energy spectrum inferred by the Stokes $V$ ZDI depends sensitively on the harmonic parameterisation of the surface magnetic field.

\begin{figure*}[!th]
\centering
\figps{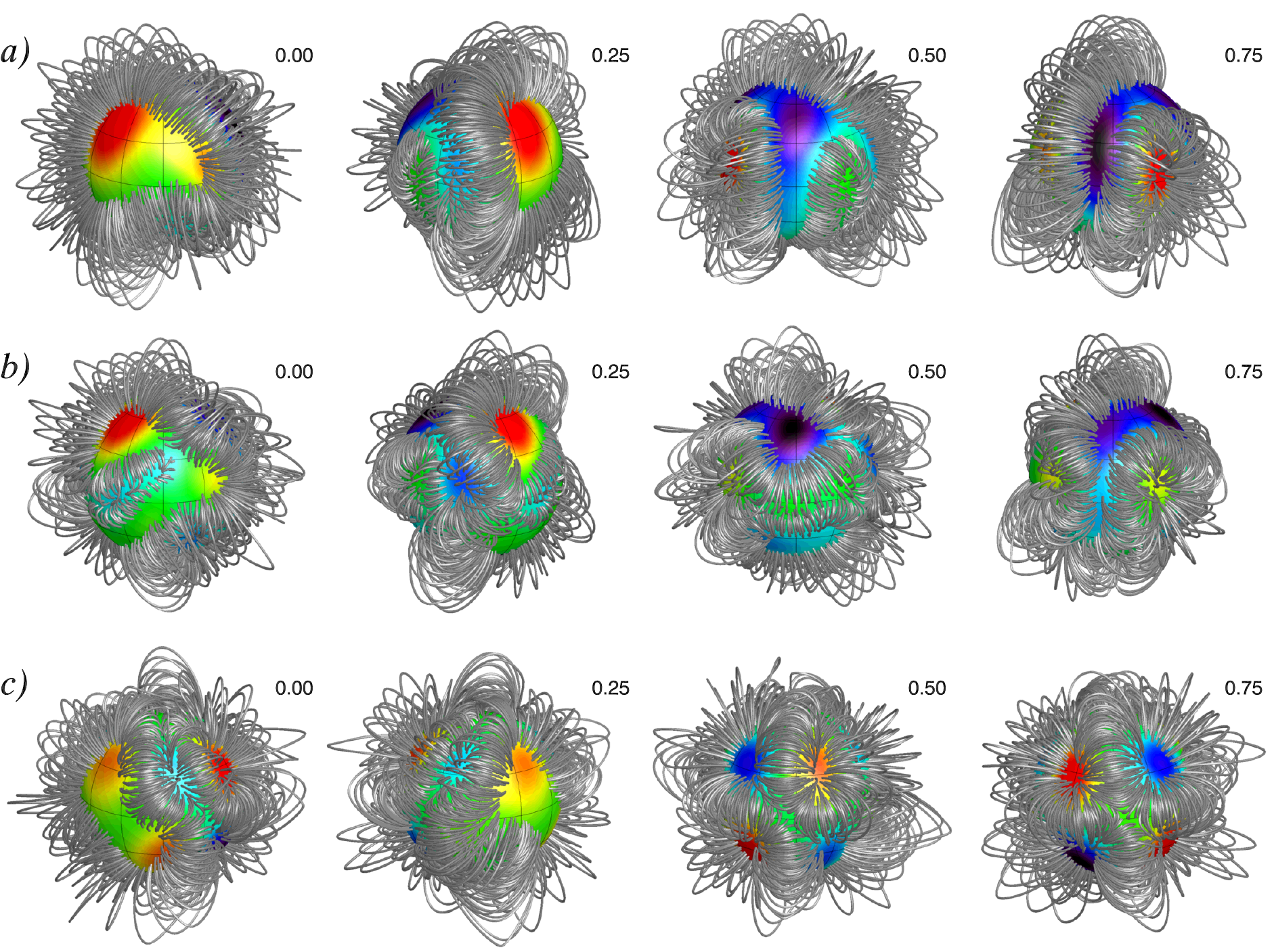}
\caption{Magnetospheric magnetic field topology of \sco\ obtained by extrapolating ZDI surface magnetic field maps. Different rows show results for {\bf a)} Model 2, {\bf b)} Model 3, and {\bf c)} Model 4. In each case the star is shown at four rotational phases as indicated above each panel.}
\label{fig:psse}
\end{figure*}

\subsection{Potential field extrapolation}

As mentioned in Sect.~\ref{pssf:intro} ZDI maps of the stellar surface magnetic fields are often used as the basis for studying the extended magnetospheric field topology with the help of the potential field extrapolation methods. Here we have applied the PSSF extrapolation technique described by \citet{jardine:2002a,jardine:2013} to the radial field component of each ZDI map that we have derived for \sco. The same source surface radius of $R=2 R_\star$ as used by D06 was adopted for all extrapolations. A 3-D rendering of the magnetic field lines for the magnetospheric field structures corresponding to different ZDI maps is presented in Fig.~\ref{fig:psse}. We omitted extrapolation for Model~1 from this figure since it is very similar to Model~2.

Analysis of Fig.~\ref{fig:psse} suggests that excluding the toroidal field from ZDI reconstruction (Model~3) has a relatively small impact on the magnetospheric field topology. However, some difference is seen for rotational phase 0.5: the field structure extrapolated from the ZDI map of Model 3 exhibits additional field loops at the disk centre.

On the other hand, the extended field topology corresponding to Model~4 is entirely different from the results obtained for Models 2 and 3. The extended field is more complex, especially at phases 0.50 and 0.75. One may expect that variability of the X-ray emission due to the material trapped in such magnetosphere will differ from the emission corresponding to the magnetospheric structures of Models 2 and 3. We emphasise again that, unlike the other two PSSF extrapolations presented in Fig.~\ref{fig:psse}, the extended field structure of Model~3 is self-consistent with the respective ZDI map.

\section{Conclusions and discussion}
\label{concl}

\begin{figure*}[!th]
\centering
\fifps{6.0cm}{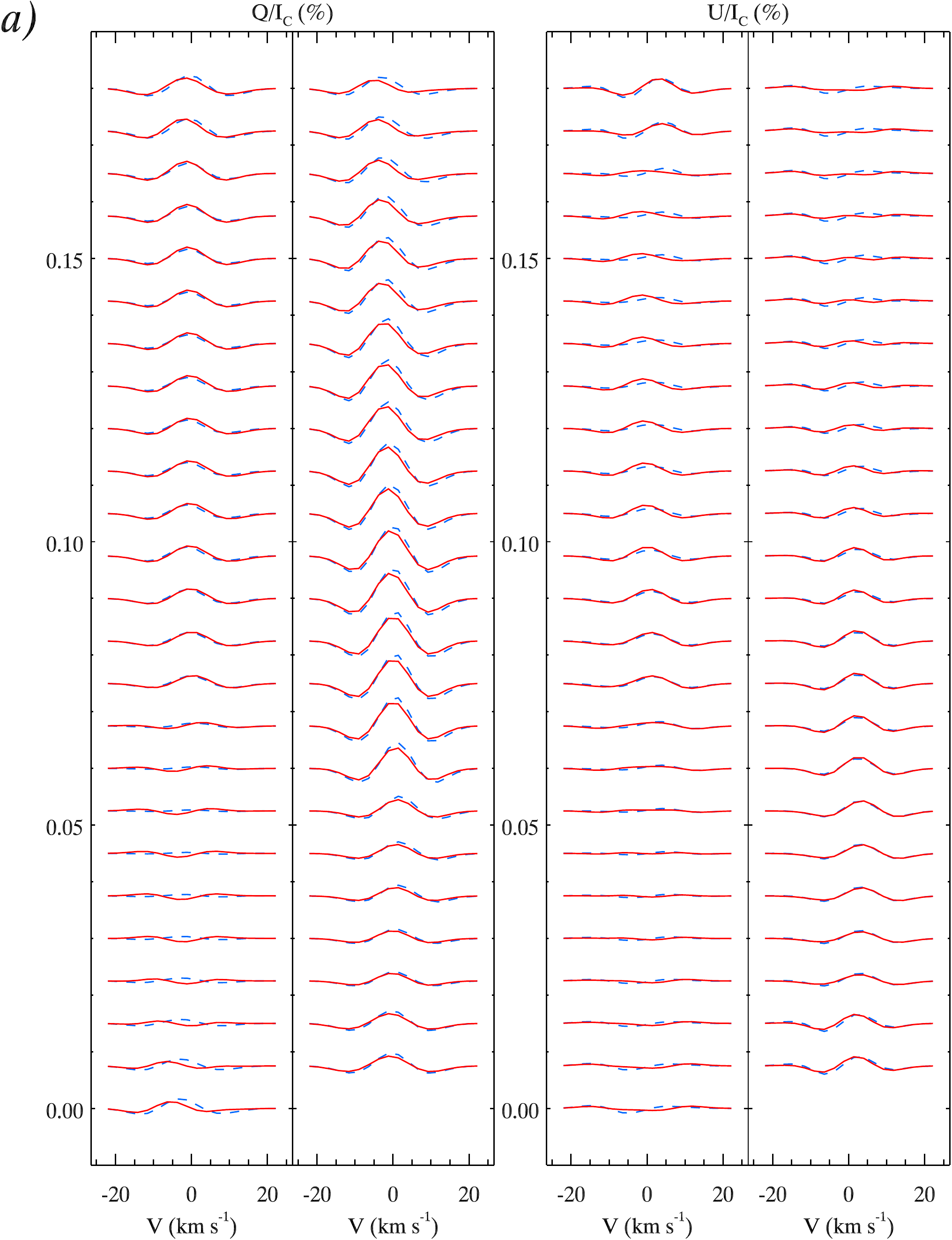}\hspace*{0.15cm}
\fifps{6.0cm}{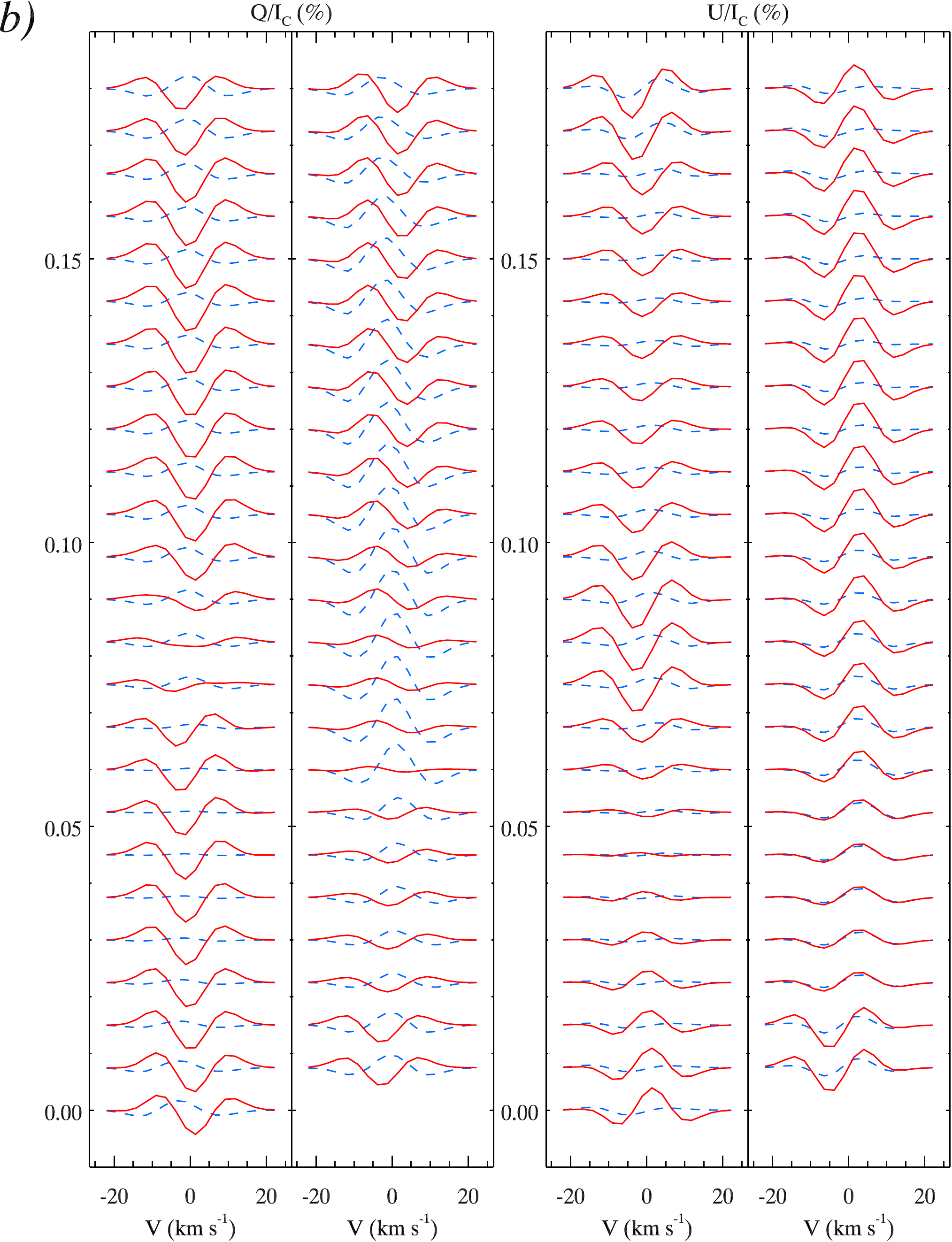}\hspace*{0.15cm}
\fifps{6.0cm}{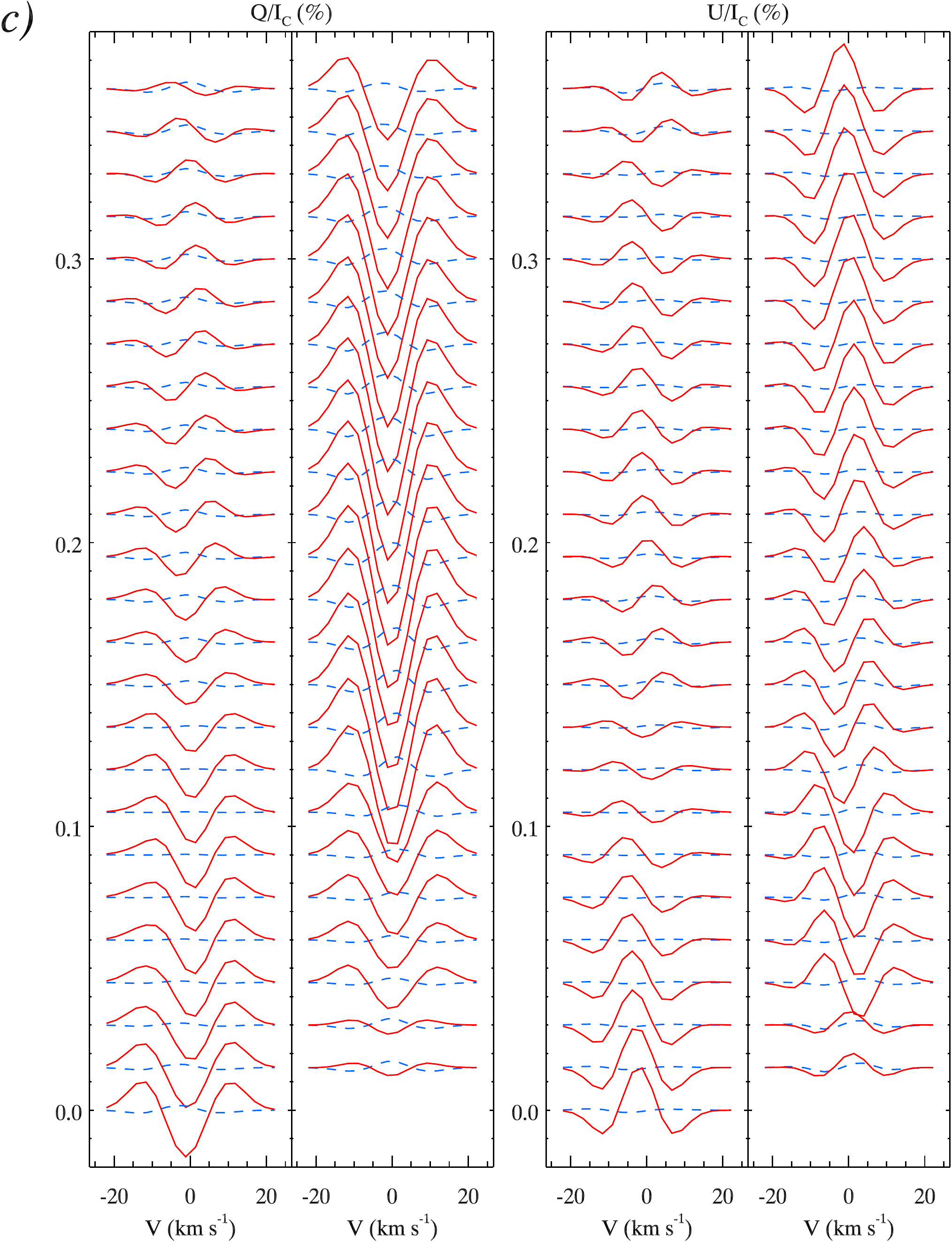}
\caption{Comparison of the synthetic LSD Stokes $QU$ profiles predicted by different ZDI inversions. In each panel the dashed line shows the linear polarisation profiles for Model 2. The solid lines correspond to {\bf a)} Model 1, {\bf b)} Model 3, and {\bf c)} Model 4. The Stokes $QU$ spectra are shown for the same set of rotational phases as the Stokes $V$ profiles in Figs.~\ref{fig:map1} and \ref{fig:map2}.}
\label{fig:prfQU}
\end{figure*}

\begin{figure}[!th]
\centering
\fifps{8cm}{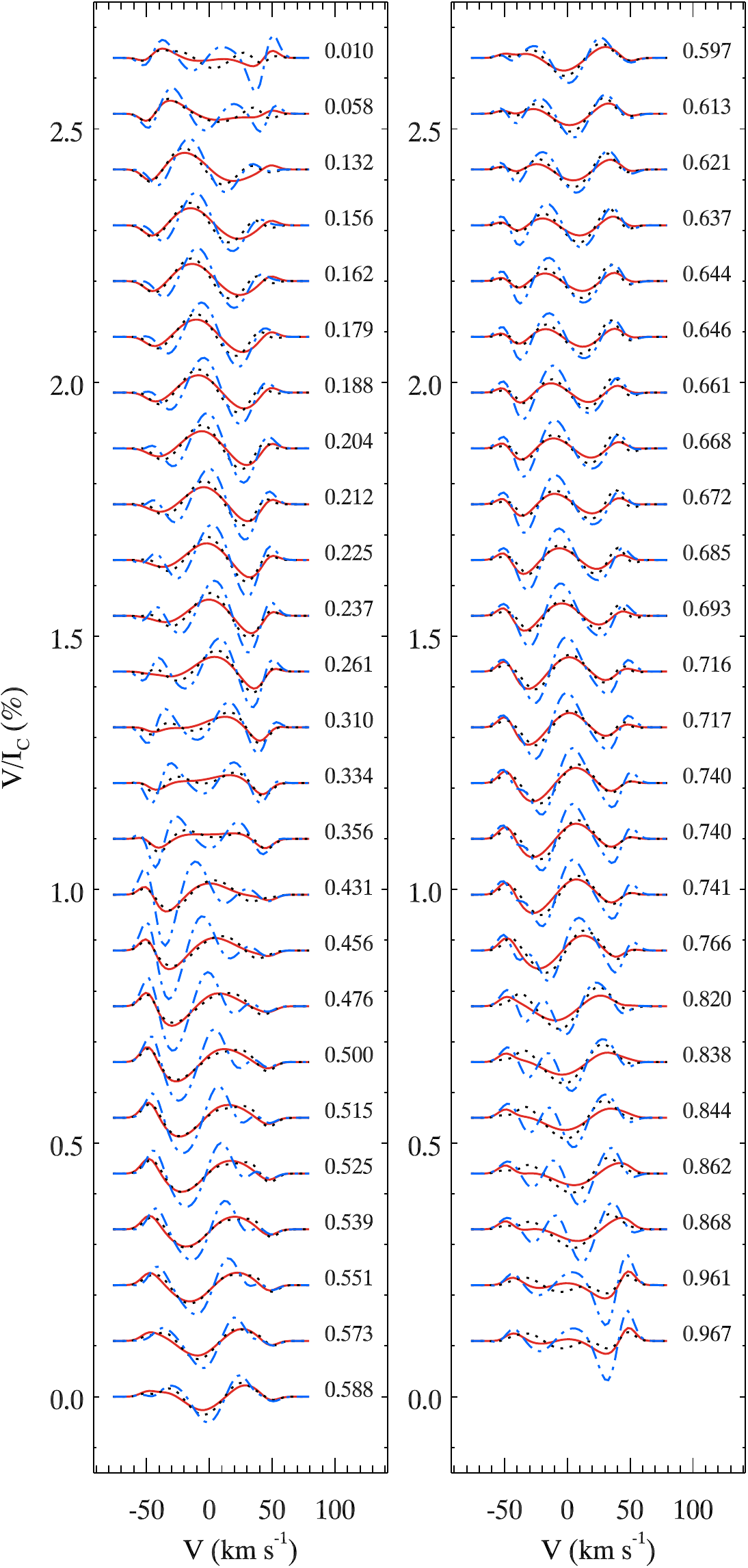}
\caption{Synthetic Stokes $V$ profiles corresponding to the magnetic field distributions of Model 2 (solid line), Model 3 (dotted line), and Model 4 (dash-dotted line) for the projected rotational velocity of \vs\,=\,50~\kms.}
\label{fig:fast}
\end{figure}

In this study we have carried out a detailed investigation of the reliability of the Stokes $V$ ZDI reconstruction of the surface magnetic field topology for the early-B star \sco. Specifically, we have fitted the observed circular polarisation profile timeseries employing different forms of the surface magnetic map parameterisation and compared the resulting surface distributions in terms of their geometry, magnetic energy and boundary condition for the potential field extrapolation. The main findings of our investigation are the following:
\begin{itemize}
\item The ZDI with general harmonic parametrisation gives a very similar result to the direct ZDI inversion constrained with the Tikhonov regularisation. Thus, for the relatively complex magnetic field topology of \sco, the harmonic parameterisation does not offer any advantages over the direct inversion in terms of overcoming biases and cross-talks typical of the Stokes $V$ ZDI reconstructions.\\
\item The magnetic field map obtained with the general harmonic ZDI inversion is fully consistent with the results published by \citet{donati:2006b} and \citet{donati:2009}. At the same time, contrary to these authors we found that the energy of the toroidal field component is comparable, if not larger, than the poloidal one.\\
\item We were able to successfully fit the available Stokes $V$ timeseries with two types of restricted harmonic fields. On the one hand, the observations can be acceptably reproduced ignoring the toroidal field altogether. On the other hand, a fit with toroidal field plus a poloidal contribution with coupled vertical and horizontal vector components, as required by the potential field extrapolation technique, is also feasible.\\
\item The ZDI models derived in our study successfully reproduce the Stokes $V$ profile variability with a static magnetic field geometry and no differential rotation. This implies that the field was stable on the time scale of five years covered by the available spectropolarimetric observations.
\end{itemize}

The three ZDI maps obtained using alternative forms of the harmonic field parameterisation are significantly different in terms of their topological details and the average field strength. They also yield somewhat different distributions of the magnetic energy as a function of the spherical-harmonic angular degree $\ell$ and different extended magnetospheric field structure according to the PSSF extrapolation method. At the same time, these field maps yield essentially indistinguishable theoretical Stokes $V$ profiles. We interpret this as another manifestation of the uniqueness problems, discussed extensively by previous ZDI studies \citep{donati:1997a,kochukhov:2002c,rosen:2012}, associated with limited information content of the Stokes $V$ timeseries.

The noticeable increase of the reconstructed field strength and complexity from Model 3 (no toroidal field) to Model 4 (coupled vertical and horizontal poloidal field components) suggests that the latter degree of freedom is more important for ZDI than the mere presence or absence of the toroidal field. While many ZDI studies pondered on the latter aspect, none has tried to physically interpret the fact that the best-fitting ZDI maps often require different magnitude and even the sign for the vertical and horizontal components of the same spherical-harmonic terms. At least for the case of stable, fossil magnetic fields of early-type stars a physical motivation for such geometries is absent. The only existing detailed theoretical models of fossil field geometries \citep{braithwaite:2006,braithwaite:2008} match the stellar field to the vacuum exterior condition, implying the same coupling of the vertical and horizontal harmonic terms as assumed in our ZDI Model 4.

It is obvious (e.g. see Table~\ref{tbl:zdi}) that attaining the same fit quality with restricted harmonic field parameterisations requires significantly stronger field than in the case of the ZDI with general spherical-harmonic expansion (Model 2). One can, therefore, insist that the latter map should be preferred based on the simplicity or minimum energy argument. However, this reasoning ignores the physical nature of the problem, which for early-type stars requires a consistent field topology all the way from the stellar surface to the circumstellar environment. This consistency is achieved by Model 4 but not by Models 2 or 3. Nevertheless, we emphasise that simplicity or physical arguments aside, the available Stokes $V$ observational data by itself definitely excludes $\ell=1$--2 models but cannot provide a distinction between the three high-order harmonic field topologies.

Extending spectropolarimetric observations to linear polarisation (Stokes $Q$ and $U$ parameters) provides a promising way to distinguish degenerate harmonic field models. Figure~\ref{fig:prfQU} compares theoretical Stokes $Q$ and $U$ timeseries for all four ZDI magnetic maps derived in our paper. It is evident that the profiles of Models 1 and 2 are still virtually indistinguishable since they correspond to essentially identical surface maps. On the other hand, the predicted $Q$ and $U$ profiles of Models 2 and 3 are different for all but a few rotational phases. Furthermore, the strong-field Model 4 yields a high-amplitude linear polarisation signal that is straightforward to distinguish from Models 2 and 3. Thus, it appears that linear spectropolarimetry is critical for solving the problem of non-uniqueness of the Stokes $V$ ZDI modelling of \sco. Unfortunately, a comparison of several Stokes $Q$ and $U$ observations of this star available in the ESPaDOnS archive with the ZDI model predictions is currently inconclusive: the observations do not have a sufficient S/N ratio to test the models using single spectral lines, and the use of higher precision Stokes $Q/U$ LSD profiles for such a test requires more sophisticated modelling \citep[see][]{rosen:2015} that is beyond the scope of this paper.

Finally, we wish to comment that the Stokes $V$ ZDI modelling of \sco\ is significantly complicated by slow rotation of this star. In this case the rotational Doppler broadening is comparable to the local line profile width and therefore the spatial Doppler resolution is ineffective compared to fast rotators which are usually targeted by DI. As demonstrated by Fig.~\ref{fig:fast}, if the projected rotational velocity of \sco\ would be several times larger than its \vs\,=\,6~\kms, one could easily discern the Stokes $V$ profiles of Model 4 from the timeseries of the two other harmonic maps. The Models 2 and 3, however, still yield nearly identical Stokes $V$ profiles for this fictitious fast-rotating \sco\ analog. 

Details of the surface magnetic field geometry of \sco, in particular the relative strengths of its poloidal and toroidal field components, have important implications for theoretical understanding of massive star magnetism. The fields in these stars are believed to be fossil remnants from an earlier evolutionary phase. Magnetohydrodynamical (MHD) simulations of the transition of an initially random fossil field to a stable configuration were carried out by \citet{braithwaite:2006} and \citet{braithwaite:2008}. It was found that, in order to maintain stability on stellar evolutionary time-scales, the interior field must have a mixed poloidal-toroidal configuration. However, these MHD models do not anticipate the presence of a significant toroidal field component at the stellar surface. In this context, confirmation of such fields at the surface of \sco\ would represent a major challenge for the current theoretical models, possibly hinting at an operation of some physical mechanism other than a simple fossil field relaxation.

To summarise our study, a Stokes $V$ ZDI inversion is significantly uncertain for a star that, like \sco, rotates slowly and exhibits a complex, non-dipolar surface magnetic field. Some key characteristics of the surface field topology (the presence of toroidal field, major geometrical features, typical field strength, degree of complexity, etc.) cannot be reliably ascertained from first principles. An assessment of these characteristics requires making a strong assumption (equivalently, choosing one of several possible surface field parameterisations) about the field without an overly compelling physical reason. Obtaining Stokes $Q$ and $U$ observations and incorporating these data in ZDI inversions represents the only generic solution to this problem.

\begin{acknowledgements}
OK acknowledges financial support from the Knut and Alice Wallenberg Foundation, the Swedish Research Council, and the G\"oran Gustafsson Foundation. GAW is supported by a Discovery Grant from the Natural Science and Engineering Research Council (NSERC) of Canada.
\end{acknowledgements}

\end{document}